\documentstyle[12pt,epsfig]{article}

\newcommand{\be}{\begin{equation}}
\newcommand{\ee}{\end{equation}}
\newcommand{\Dlt}{\Delta}

\newcommand{\prt}{\partial}
\newcommand{\br}{{\bf r}}
\newcommand{\bk}{{\bf k}}
\newcommand{\bt}{\beta}

\newcommand{\ep}{\varepsilon}

\newcommand{\ra}{\rightarrow}
\newcommand{\sgm}{\sigma}

\newcommand{\om}{\omega}

\newcommand{\dgr}{\dagger}

\begin{document}

\begin{center}
{\Large {\bf Normal and Anomalous Averages for Systems with 
Bose-Einstein Condensate} \\ [5mm]

V.I. Yukalov$^{1,2}$ and E.P. Yukalova$^{1,3}$} \\ [3mm]

{\it $^1$Institut f\"ur Theoretische Physik, \\
Freie Universit\"at Berlin, Arnimallee 14, D-14195 Berlin, Germany \\ [2mm]

$^2$Bogolubov Laboratory of Theoretical Physics, \\
Joint Institute for Nuclear Research, Dubna 141980, Russia \\ [2mm]

$^3$Department of Computational Physics, Laboratory of 
Information Technologies, \\
Joint Institute for Nuclear Research, Dubna 141980, Russia}

\end{center}

\vskip 2cm

\begin{abstract}

The comparative behaviour of normal and anomalous averages as functions 
of momentum or energy, at different temperatures, is analysed for systems 
with Bose-Einstein condensate. Three qualitatively distinct temperature
regions are revealed: The critical region, where the absolute value of the 
anomalous average, for the main energy range, is much smaller than the 
normal average. The region of intermediate temperatures, where the absolute 
values of the anomalous and normal averages are of the same order. And the 
region of low temperatures, where the absolute value of the anomalous average,
for practically all energies, becomes much larger than the normal average. 
This shows the importance of the anomalous averages for the intermediate
and, especially, for low temperatures, where these anomalous averages
cannot be neglected.

\end{abstract}

\vskip 1cm


\vskip 2cm


\newpage

\section{Introduction}

In the last years, the physics of dilute Bose gases has received much 
consideration both in theoretical and experimental aspects. An exhaustive 
list of references can be found in the recent review articles [1--4] and 
books [5,6]. The theoretical description of dilute systems with Bose-Einstein 
condensate is, to a large extent, based on the ideas of the Bogolubov theory, 
advanced in the papers [7,8] and thoroughly expounded in books [9,10]. This 
is because the present-day experiments deal in the majority of cases exactly 
with such dilute Bose gases.

The starting idea of the Bogolubov theory is the breaking of gauge symmetry 
by means of the so-called Bogolubov shift for the field operator
$$
\psi(\br) =\eta(\br) +\psi_1(\br) \; ,
$$
where $\eta(\br)$ is the condensate wave function, related to a coherent 
state [11,12] and defining the condensate density $\rho_0(\br)=|\eta(\br)|^2$.
As soon as the gauge symmetry is broken, in addition to the normal averages 
$<\psi^\dgr(\br)\psi(\br')>$ or $<a_k^\dgr a_k>$, depending on whether the 
real or momentum space is considered, the anomalous averages 
$<\psi(\br)\psi(\br')>$ or $<a_k a_{-k}>$ arise in the theory. 

The physical meaning and general features of the normal averages, which 
are directly connected with the reduced density matrices [13], seem to 
be familiar and clear to everyone. Contrary to this, the role and general 
behaviour of the anomalous averages are not always properly understood.
There exists a very widespread delusion that the anomalous averages can
be omitted at all, even at low temperatures. One often ascribes this
unjustified trick to Popov, calling it the "Popov approximation". 
However, as is easy to infer from the original works by Popov [14--17], 
he has never done anything like that. He considered the temperatures in 
the close vicinity of the critical temperature $T_c$. When the temperature
$T$ tends to $T_c$, then the condensate density $\rho_0$ tends to zero. 
The anomalous averages are proportional to $\rho_0$, hence, also tend to 
zero together with $\rho_0$. Contrary to this, the normal averages are
proportional to the density of noncondensed particles $\rho_1$, which is 
close to the total density $\rho$, when $T\approx T_c$. That is why in a
narrow neighbourhood of $T_c$ the anomalous averages automatically become
much smaller than the normal averages, without any additional assumptions.
However, for a dilute gas at low temperatures $T\ll T_c$, the condensate 
density can become comparable with the total density, $\rho_0\approx\rho$,
while, vice versa, the density of noncondensed particles can be much 
smaller than $\rho$. Then anomalous averages can be much larger than the 
normal ones, and there is no any grounds for omitting the former. Popov 
[14--17] has never neglected the anomalous averages at low temperatures 
$T\ll T_c$.

It is the aim of the present paper to analyse the behaviour of the 
anomalous averages at different temperatures and to illustrate the 
properties of the anomalous averages as compared to those of the normal 
averages. Such a comparison reveals the role of the anomalous averages 
emphasizing the importance of their contribution, especially at low 
temperatures, where the anomalous averages can never be neglected.

Throughout the paper, the system of units is employed, where the Planck
constant $\hbar=1$ and the Boltzmann constant $k_B=1$ are set to unity.

\section{Normal and Anomalous Averages}

The behaviour of the averages can be rather different at varying temperature.
It is possible to distinguish three principally different regions. For this
purpose, the temperature $T$ should be compared with the characteristic value 
$\rho\Phi_0$, where $\rho\equiv N/V$ is the particle density for $N$ particles
in volume $V$ and
\be
\label{1}
\Phi_0 \equiv \int \Phi(\br)\; d\br
\ee
is the integral of the interaction potential $\Phi(\br)$, assumed to be 
integrable. The value $\rho\Phi_0$ characterizes the average strength of 
particle interactions. Another typical quantity is the critical temperature
$T_c$, which for a weakly interacting gas is close to the critical 
temperature for an ideal gas
\be
\label{2}
T_c = \frac{2\pi}{m}\left [ \frac{\rho}{\zeta(3/2)}\right ]^{2/3} \; ,
\ee
where $m$ is particle mass and $\zeta(3/2)=2.612$. The relation between 
$\rho\Phi_0$ and $T_c$ can be obtained by introducing the scattering 
length $a_s$ through the equation
\be
\label{3}
\Phi_0 \equiv 4\pi \; \frac{a_s}{m} \; .
\ee
For the repulsive interactions, the scattering length is positive, which 
is assumed in what follows. Then we find
$$
\frac{\rho\Phi_0}{T_c} =  2\left [ \zeta\left ( \frac{3}{2}\right )
\right ]^{2/3} \rho^{1/3} a_s \; .
$$
Keeping in mind a dilute gas, for which by definition
\be
\label{4}
\rho a_s^3 \ll 1 \; ,
\ee
we obtain
\be
\label{5}
\frac{\rho\Phi_0}{T_c}\ll 1 \; .
\ee

In this way, we can distinguish the following temperature regions. The
interval of low temperatures is
\be
\label{6}
0 \leq T \leq \rho \Phi_0 \; .
\ee
Intermediate temperatures correspond to the inequalities
\be
\label{7}
\rho \Phi_0 < T \ll T_c \; .
\ee
And the temperatures $T\sim T_c$ close to the critical point pertain to 
the critical temperature region.

The critical region of temperatures in the close vicinity of the 
condensation temperature $T_c$ was analysed by Popov [14--16]. In this 
region, the anomalous averages turned out to be small as compared to the
normal ones. This fact is very simple to understand, since when temperature
tends to $T_c$, the anomalous averages are proportional to the density of
condensed particles $\rho_0$, which tends to zero as $T\ra T_c$. But the
normal averages are proportional to the density of noncondensed particles 
$\rho_1$, which tends, as $T\ra T_c$, to the finite value $\rho$ of the
total particle density. This is why in the close neighbourhood of $T_c$ 
the contribution of the anomalous averages is small as compared to that 
of the normal averages.

However at the temperatures outside the critical region, the situation can 
be drastically different. And we have to study these lower temperatures more 
attentively. We shall consider a uniform dilute gas, for which the inequality 
(4) is valid.

In a dilute gas, for $T\ll T_c$, almost all particles are condensed. 
Then the Bogolubov theory [7--10] is applicable. Following this theory, 
one makes the Bogolubov shift of the field operators and retains quadratic 
fluctuations of the noncondensed particles, which results in the Hamiltonian
\be
\label{8}
H = \frac{1}{2}\; N_0\rho_0\Phi_0  - \mu N_0 + \sum_{k\neq 0}
\om_k a_k^\dgr a_k + \frac{1}{2}\; \sum_{k \neq 0} \Dlt_k \left (
a_k^\dgr a_{-k}^\dgr + a_{-k} a_k \right ) \; ,
\ee
in which $N_0$ is the number of condensed particles, $\rho_0\equiv N_0/V$
is their density,
\be
\label{9}
\om_k \equiv \frac{k^2}{2m} +\rho_0 (\Phi_0 + \Phi_k) - \mu \; ,
\ee
$\mu$ is chemical potential, and
\be
\label{10}
\Dlt_k \equiv \rho_0 \Phi_k \; ,
\ee
with $\Phi_k$ being the Fourier transform of the interaction potential,
$$
\Phi_k = \int \Phi(\br) e^{-i\bk\cdot\br} \; d\br \; .
$$
Under the considered conditions, $N_0\approx N$ is close to the total 
number of particles $N$.

The Hamiltonian (8) is diagonalized by means of the Bogolubov canonical 
transformation
$$
a_k = u_k b_k + v_{-k}^* b_{-k}^\dgr \; ,
$$
in which the coefficient functions can be chosen to be symmetric and real, 
such that $u_k^*=u_{-k}=u_k$ and  $v_k^*=v_{-k}=v_k$. They are defined by 
the relations
$$
u_k^2 -v_k^2 = 1 \; , \qquad u_kv_k = -\; \frac{\Dlt_k}{2\ep_k} \; ,
$$
$$
u_k^2 = \frac{\sqrt{\ep_k^2+\Dlt^2_k}+\ep_k}{2\ep_k} =
\frac{\om_k+\ep_k}{2\ep_k} \; , \qquad
v_k^2 = \frac{\sqrt{\ep_k^2+\Dlt^2_k}-\ep_k}{2\ep_k} =
\frac{\om_k-\ep_k}{2\ep_k} \; ,
$$
where the Bogolubov spectrum is
\be
\label{11}
\ep_k =\sqrt{\om_k^2-\Dlt^2_k} \; .
\ee
Under this transformation, Hamiltonian (8) is recast to the diagonal form
\be
\label{12}
H_B = E_0 + \sum_{k\neq 0} \ep_k b_k^\dgr b_k - \mu N_0 \; ,
\ee
with the ground-state energy
$$
E_0 =\frac{1}{2}\; N_0\rho_0 \Phi_0  + \frac{1}{2}\; 
\sum_{k\neq 0} (\ep_k - \om_k) \; .
$$

The separation of the condensate contribution, as is well known, is 
meaningful only when the momentum distribution of particles becomes 
singular at the point $k=0$, which is associated with the particle 
spectrum touching zero at $k=0$. Only then there is a reason of separating 
the zero-momentum terms [18], which is equivalent to the Bogolubov shift 
of the field operators [9,10]. At the same time, this guarantees a gapless 
spectrum, in agreement with the theorems by Hugenholtz and Pines [19] and 
Bogolubov [9,10]. These requirements can be simply formulated as the 
condition
\be
\label{13}
\lim_{k\ra 0} \ep_k = 0 \; , \qquad \ep_k \geq 0 \; .
\ee
Then from spectrum (11), we get the chemical potential
\be
\label{14}
\mu=\rho_0 \Phi_0 \; .
\ee
Another way of defining the chemical potential is by minimizing the
thermodynamic potential
$$
\Omega = - T\ln{\rm Tr}\; e^{-\bt H}
$$
with respect to $N_0$. Then from the equation 
$$
\frac{\prt\Omega}{\prt N_0} = \; < \frac{\prt H}{\prt N_0} > \; 
= \; 0 \; ,
$$
in the frame of the Bogolubov approximation, one obtains the same chemical 
potential (14). Respectively, Eq. (11) acquires the known form of the 
Bogolubov spectrum
\be
\label{15}
\ep_k = \sqrt{\frac{\Dlt_k}{m}\; k^2 +\left (
\frac{k^2}{2m}\right )^2 } \; .
\ee

The diagonal Hamiltonian (12) makes it possible to find explicit 
expressions for different averages [20--22]. Our concern here is the 
normal average
\be
\label{16}
n_k \equiv \; <a_k^\dgr a_k>
\ee
and the anomalous average
\be
\label{17}
\sgm_k\equiv \; < a_k a_{-k}> \; ,
\ee
whose absolute values are to be compared.

First of all, we can derive a general relation between averages (16) and
(17), following from the Bogolubov inequality
$$
\left | <\hat A\hat B>\right |^2 \leq \left (
<\hat A\hat A^+><\hat B^+\hat B> \right ) \; ,
$$
valid for the averages involving any two operators. Setting here $\hat 
A=a_k$ and $\hat B=a_{-k}$, we find
$$
|<a_ka_{-k}>|^2 \leq \left ( <a_k a_k^\dgr><a_{-k}^\dgr a_{-k}>
\right ) \; .
$$
This yields
\be
\label{18}
\sgm_k^2 \leq n_k (1 + n_k) \; .
\ee

Calculating expressions (16) and (17), we have the normal average
\be
\label{19}
n_k = \frac{\sqrt{\ep_k^2+\Dlt^2_k}}{2\ep_k}\; {\rm coth}\left (
\frac{\ep_k}{2T}\right ) -\; \frac{1}{2}
\ee
and the anomalous average
\be
\label{20}
\sgm_k = -\; \frac{\Dlt_k}{2\ep_k}\; {\rm coth}\left (
\frac{\ep_k}{2T}\right ) \; .
\ee
Inequality (18) holds true, since
$$
n_k ( 1+ n_k) - \sgm_k^2 = \left [ 2 {\rm sinh}\left ( 
\frac{\ep_k}{2T}\right ) \right ]^{-2} \; .
$$

The spectrum (15) possesses the asymptotic properties
$$
\ep_k \simeq ck \qquad (k\ra 0) \; , 
$$
\be
\label{21}
\ep_k \simeq \frac{k^2}{2m} \qquad (k\ra \infty) \; ,
\ee
where $c\equiv\sqrt{\rho_0\Phi_0/m}$ is the sound velocity, which tells us
that $\ep_k$ varies in the range
$$
0\leq \ep_k < \infty \; .
$$
It is, therefore, convenient to analyse the behaviour of Eqs. (19) and 
(20) with respect to the variable $\ep_k$.

At small excitation energy, such that
$$
\ep_k \ll \Dlt_k \; , \qquad \ep_k \ll T \; ,
$$
the normal average (19) has the asymptotic behaviour
\be
\label{22}
n_k \simeq \frac{T\Dlt_k}{\ep_k^2} + \frac{\Dlt_k}{12T} +
\frac{T}{2\Dlt_k} - \; \frac{1}{2} + \left (
\frac{\Dlt_k}{3T} - \; \frac{T}{\Dlt_k} - \; \frac{\Dlt_k^3}{90T^3} 
\right ) \frac{\ep_k^2}{8\Dlt_k^2} \; ,
\ee
and the anomalous average (20) is
\be
\label{23}
\sgm_k \simeq - \; \frac{T\Dlt_k}{\ep_k^2} - \; \frac{\Dlt_k}{12T} +
\frac{\Dlt_k\ep_k^2}{720T^3} \; .
\ee
Hence at low energies $\ep_k$, or low momenta $k$, the values $n_k$ and 
$|\sgm_k|$ asymptotically coincide for all temperatures. The region of 
small momenta $k$ usually gives the largest contribution to the integrals
representing the observable quantities. Consequently, the anomalous average 
plays an important role.

At large excitation energy, such that
$$
\ep_k \gg \Dlt_k \; , \qquad \ep_k \gg T \; ,
$$
the normal average behaves as
\be
\label{24}
n_k \simeq \left (\frac{\Dlt_k}{2\ep_k} \right )^2 -
\left (\frac{\Dlt_k}{2\ep_k} \right )^4 + e^{-\bt\ep_k} \; ,
\ee
while the anomalous average, as
\be
\label{25}
\sgm_k \simeq - \; \frac{\Dlt_k}{2\ep_k} \left ( 1 + 2 e^{-\bt\ep_k}
\right ) \; .
\ee 
This shows that for $k\ra\infty$, the absolute value of the anomalous average 
is always much larger than $n_k\ll|\sgm_k|$.

From Eqs. (22) and (23), we have
\be
\label{26}
\lim_{k\ra 0} \left (|\sgm_k| - n_k\right ) =
\frac{1}{2}\left ( 1 -\; \frac{T}{\rho\Phi_0} \right ) \; ,
\ee
where we take into account that for the considered dilute gas 
$\rho_0\approx\rho$. Equality (26) demonstrates that, when $T>\rho\Phi_0$, 
then the value $|\sgm_k|$ is smaller than $n_k$ at $k=0$, though becomes 
larger than the latter for $k\ra\infty$, according to Eqs. (24) and (25).
This defines the region of intermediate temperatures (7), where $|\sgm_k|$
is comparable with $n_k$, being below $n_k$ for small $k$, but surpassing
$n_k$ for large $k$. And for low temperatures from the interval (6), the 
absolute value of the anomalous average $|\sgm_k|$ is larger than $n_k$ in 
the whole range of the momenta $k\geq 0$.

To illustrate pictorially the relative behaviour of the normal and anomalous 
averages, we calculate Eqs. (19) and (20) numerically. It is convenient to 
measure the temperature $T$ in units of $\rho\Phi_0$ and to treat $n_k$ and
$\sgm_k$ as the functions of the dimensionless variable
\be
\label{27}
E \equiv \frac{\ep_k}{\Dlt_k} \; .
\ee
Then Eq. (19) can be rewritten as
\be
\label{28}
n(E) =\frac{\sqrt{1+E}}{2E}\; {\rm coth}\left ( \frac{E}{2T}\right ) -\;
\frac{1}{2}
\ee
and Eq. (20) takes the form
\be
\label{29}
\sgm(E)= -\; \frac{1}{2E}\;  {\rm coth}\left ( \frac{E}{2T}\right ) \; .
\ee
The normal average (28) is non-negative for all $E\geq 0$, while the 
anomalous average (29) is always negative. To compare their absolute 
values, it is useful to introduce the notation
\be
\label{30}
A(E) \equiv |\sgm(E)|
\ee
for the absolute value of the anomalous average (29).

Figure 1 shows the typical behaviour of the anomalous value (30) and the 
normal average (28) for the low-temperature region (6), where $A(E)$ is 
always larger than $n(E)$. Figures 2 and 3 demonstrate the changes in these
functions, when the temperature rises to the intermediate region (7), where 
the averages are comparable with each other, so that $A(E)<n(E)$ at small
$E$, but at large $E$, this inequality inverts to $n(E)<A(E)$.

It is very instructive to picture the ratio $n(E)/A(E)$, clearly showing 
how much different is the normal average $n(E)$, as compared to the anomalous 
value $A(E)$. Figure 4 illustrates this for the low-temperature region (6). 
As is evident, the anomalous average essentially prevails over the normal 
one. Figure 5 corresponds to the borderline between the regions of low 
and intermediate temperatures, where the anomalous average is yet dominant. 
Finally, Fig. 6 displays the ratio $n(E)/A(E)$ for the 
intermediate-temperature region (7), when the inequality $n(E)\geq A(E)$ 
for small $E$ overturns to $n(E)<A(E)$ for large $E$.

\section{Conclusion}

In the theory with broken gauge symmetry, anomalous averages play an 
important role. Although in the critical region, close to the condensate
temperature, they become much smaller by the absolute value, as compared 
to the normal averages, however at lower temperatures, the situation turns 
to be essentially different. At temperatures $\rho\Phi_0<T\ll T_c$ in the 
dilute Bose gas, the absolute value of the anomalous average is of the same 
order as the normal average. For very low temperatures $T\leq\rho\Phi_0$, the 
anomalous average surpasses by its absolute value the normal average and can 
even be much larger than the latter. This analysis shows that, as soon as 
the gauge symmetry is broken, the anomalous averages come into play, being 
at low temperatures as important as the normal ones. It is, therefore,
principally incorrect to omit the anomalous averages for $T\ll T_c$.

Note also the fundamental importance of the anomalous averages in the 
calculations of the particle fluctuations [23--25]. In the frame of
the Bogolubov theory [7--10], the dispersion of the number-of-particle
operator $\hat N$ is
$$
\Dlt^2(\hat N) = N [ 1 + 2\lim_{k\ra 0} ( n_k +\sgm_k)] \; .
$$
The long-wave limits of $n_k$ and $\sgm_k$ separately are divergent, as 
follows from Eqs. (22) and (23). However, their singularities are the 
same in the absolute values while opposite in signs, because of which 
they cancel each other, yielding the finite limit
$$
\lim_{k\ra 0} (n_k + \sgm_k) = \frac{1}{2}\left (
\frac{T}{mc^2}\; - \; 1 \right ) \; .
$$
As a result, the dispersion of $\hat N$ becomes
$$
\Dlt^2(\hat N) = \frac{TN}{mc^2} \; ,
$$
describing normal particle fluctuations. If one would omit $\sgm_k$ in
$\Dlt^2(\hat N)$, this would lead to a divergent dispersion, hence, to
a divergent isothermal compressibility $\kappa_T=\Dlt^2(\hat N)/\rho TN$, 
which would mean the system instability [23--25]. Thus, neglecting the 
anomalous averages not merely can drastically distort quantitative values 
of observables, but also provokes qualitatively wrong unphysical 
consequences.

\vskip 5mm

{\bf Acknowledgement}

\vskip 2mm

Financial support from the German Research Foundation is acknowledged.
One of us (V.I.Y.) appreciates the Mercator Professorship of the German 
Research Foundation.

\newpage

\begin{figure}[h]
\centerline{\psfig{file=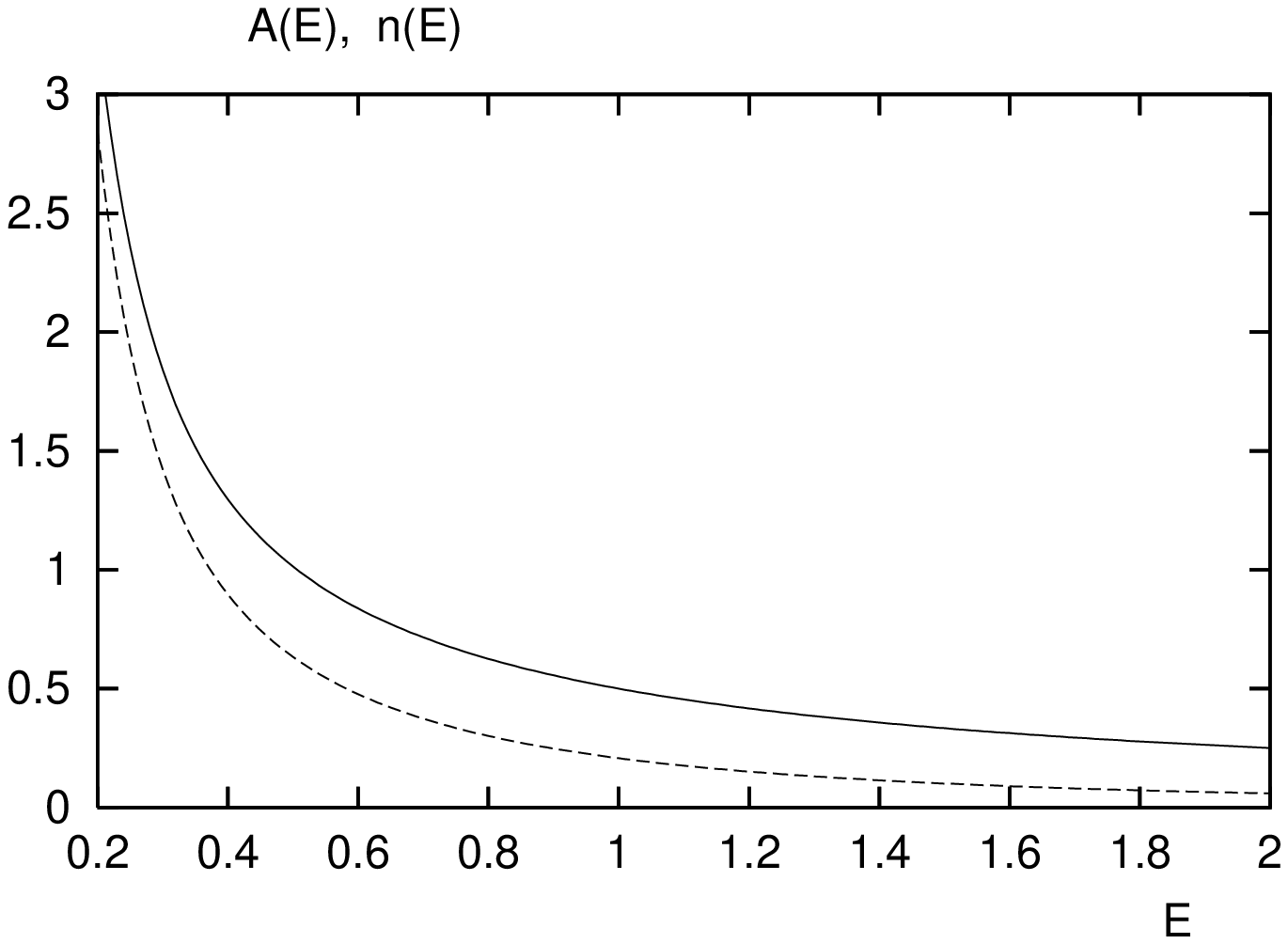,angle=270}}
\vskip 1cm
\caption{The absolute value of the anomalous average $A(E)$ (solid line) 
and the normal average $n(E)$ (dashed line) as functions of the dimensionless 
energy (27) for the reduced temperature $T/\rho\Phi_0=0.1$}
\label{fig:Fig.1} 
\end{figure}

\newpage

\begin{figure}[h]
\centerline{\psfig{file=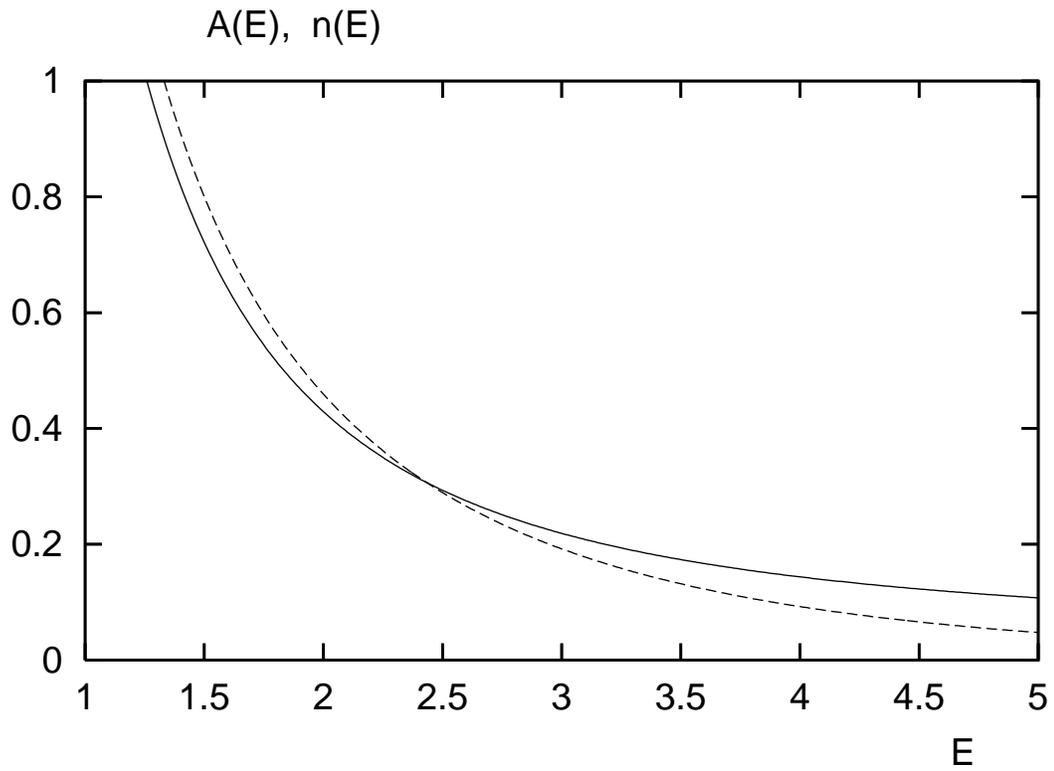,angle=270}}
\vskip 1cm
\caption{The anomalous value $A(E)$ (solid line) and the normal average 
$n(E)$ (dashed line) as functions of $E$ for the reduced temperature 
$T/\rho\Phi_0=1.5$.}
\label{fig:Fig.2} 
\end{figure}

\newpage

\begin{figure}[h]
\centerline{\psfig{file=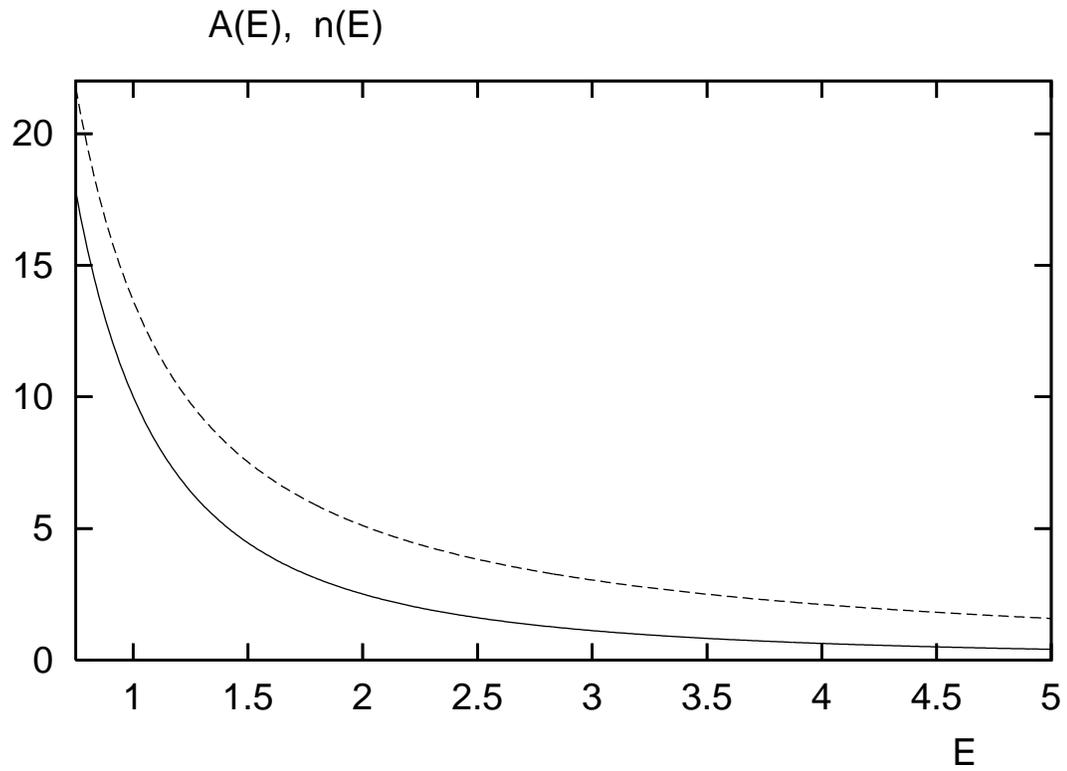,angle=270}}
\vskip 1cm
\caption{The same as in Fig. 2 for the reduced temperature 
$T/\rho\Phi_0=10$.}
\label{fig:Fig.3}
\end{figure}

\newpage

\begin{figure}[h]
\centerline{\psfig{file=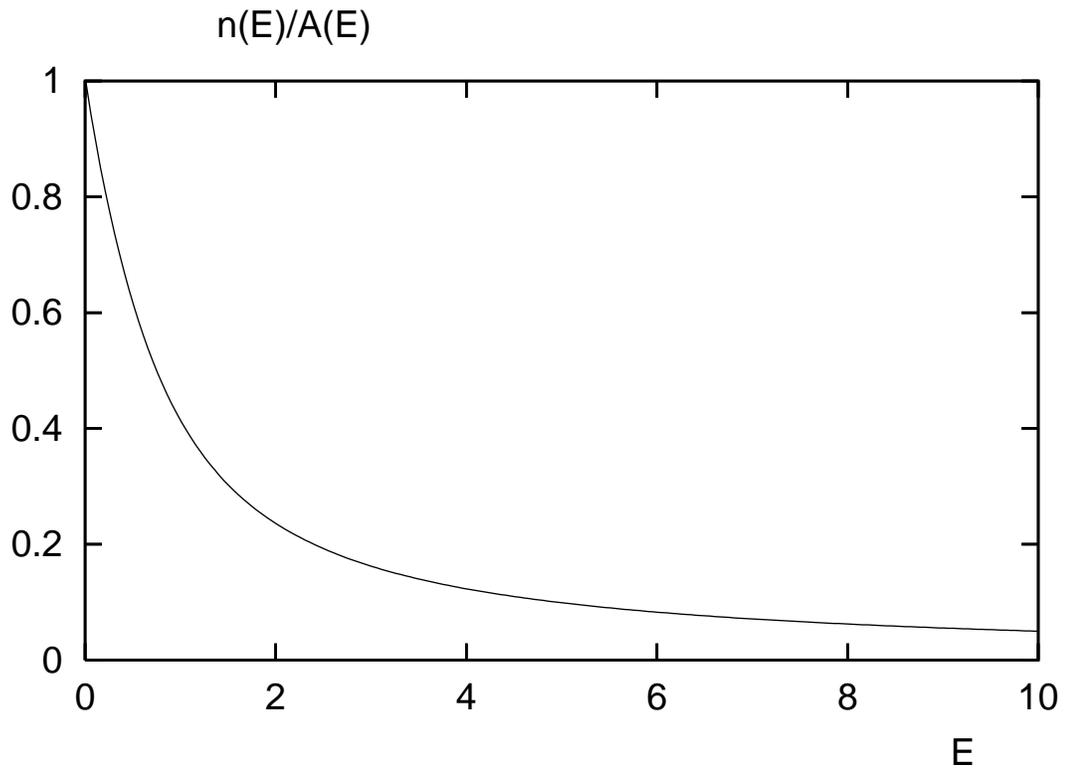,angle=270}}
\vskip 1cm
\caption{The ratio $n(E)/A(E)$ of the normal average to the absolute value 
of the anomalous average as a function of the dimensionless energy (27) 
for the low temperature $T/\rho\Phi_0=0.01$.}
\label{fig:Fig.4}
\end{figure}

\newpage

\begin{figure}[h]
\centerline{\psfig{file=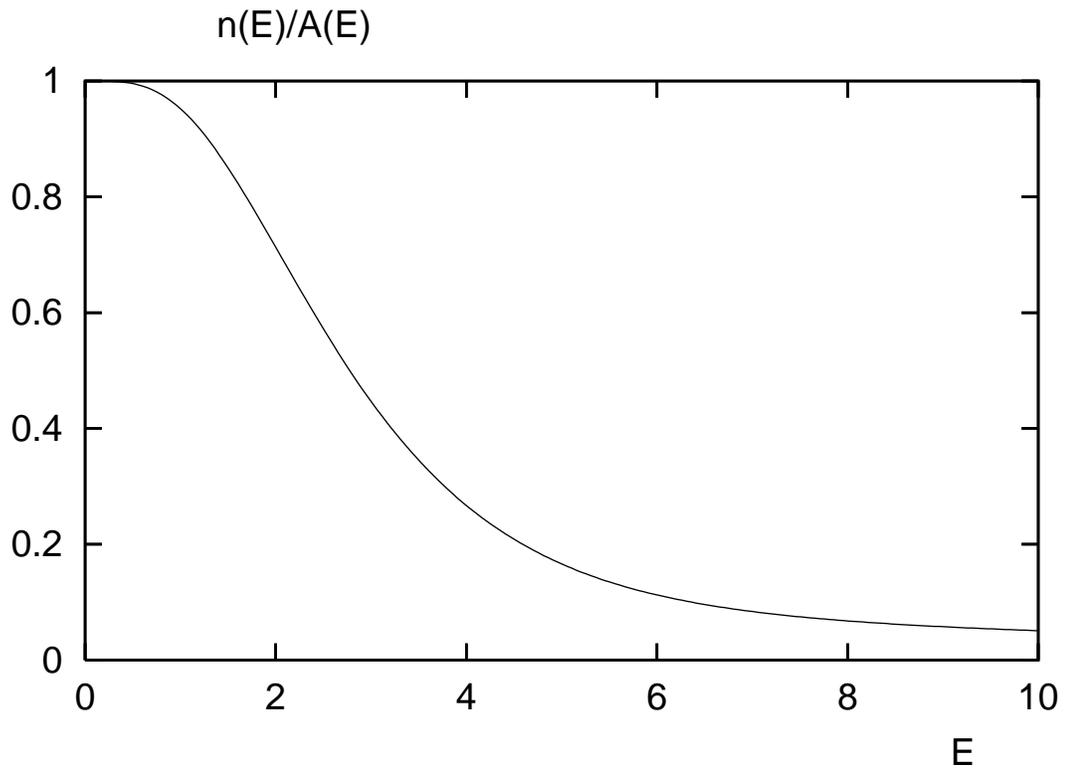,angle=270}}
\vskip 1cm
\caption{The ratio $n(E)/A(E)$ for the borderline temperature
$T/\rho\Phi_0=1$.}
\label{fig:Fig.5}
\end{figure}

\newpage

\begin{figure}[h]
\centerline{\psfig{file=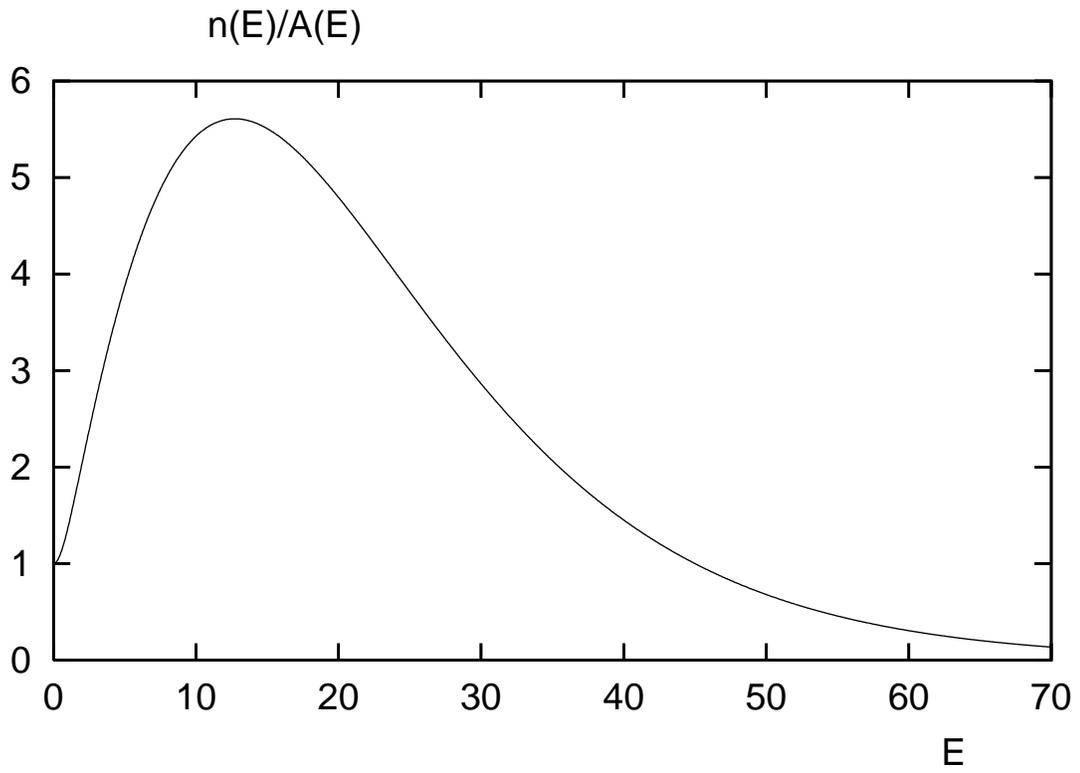,angle=270}}
\vskip 1cm
\caption{The ratio $n(E)/A(E)$ for the intermediate temperature
$T/\rho\Phi_0=10$.}
\label{fig:Fig.6}
\end{figure}

\newpage

\begin{center}
{\large{\bf Figure captions}}
\end{center}

{\bf Fig. 1}. The absolute value of the anomalous average $A(E)$ (solid 
line) and the normal average $n(E)$ (dashed line) as functions of the 
dimensionless energy (27) for the reduced temperature $T/\rho\Phi_0=0.1$.

\vskip 5mm

{\bf Fig. 2}. The anomalous value $A(E)$ (solid line) and the normal 
average $n(E)$ (dashed line) as functions of $E$ for the reduced 
temperature 
$T/\rho\Phi_0=1.5$.

\vskip 5mm 

{\bf Fig. 3}. The same as in Fig. 2 for the reduced temperature 
$T/\rho\Phi_0=10$.

\vskip 5mm 

{\bf Fig. 4}. The ratio $n(E)/A(E)$ of the normal average to the absolute 
value of the anomalous average as a function of the dimensionless energy 
(27) for the low temperature $T/\rho\Phi_0=0.01$.

\vskip 5mm 

{\bf Fig. 5}. The ratio $n(E)/A(E)$ for the borderline temperature 
$T/\rho\Phi_0=1$.

\vskip 5mm

{\bf Fig. 6}. The ratio $n(E)/A(E)$ for the intermediate temperature
$T/\rho\Phi_0=10$.

\end{document}